\documentclass[12pt]{article}

\usepackage{graphicx}

\usepackage{floatrow}

\newfloatcommand{capbtabbox}{table}[][\FBwidth]

\usepackage{multicol}

\makeatletter
\newenvironment{tablehere}
  {\def\@captype{table}}
  {}

\makeatother

\input babarsym

\def\CP{\ensuremath{C\!P}\xspace}

\def\Babar{\slshape B\kern-0.1em{\footnotesize A}\kern-0.1em B\kern-0.10em{\footnotesize A\kern-0.20em R}} 
\def\babar{\slshape B\kern-0.1em{\scriptsize A}\kern-0.1em B\kern-0.10em{\scriptsize A\kern-0.20em R}} 
\def\Bbar{\kern 0.18em\overline{\kern -0.18em B}{}} 
\def\BbarSubscr{\kern 0.25em\overline{\kern -0.18em B}{}} 
\def\Dbar{\kern 0.20em\overline{\kern -0.20em D}{}}

\def\gev{\mathrm{\,Ge\kern -0.1em V}} 
\def\mev{\mathrm{\,Me\kern -0.1em V}} 
\def\gevc{\mathrm{\,Ge\kern -0.1em V\!/}c} 
\def\gevcc{\mathrm{\,Ge\kern -0.1em V\!/}c^2} 
\def\mevcc{\mathrm{\,Me\kern -0.1em V\!/}c^2} 
\def\gevccsq{\mathrm{\,Ge\kern -0.1em V^2\!/}c^4} 
\def\ubar{\kern 0.10em \overline{\kern-0.10em u}\kern 0.05em{}} 
\def\dbar{\kern 0.15em \overline{\kern-0.15em d}\kern 0.05em{}} 
\def\sbar{\kern 0.10em \overline{\kern-0.10em s}\kern 0.05em{}} 
\def\cbar{\kern 0.10em \overline{\kern-0.10em c}\kern 0.05em{}} 
\def\bbar{\kern 0.10em \overline{\kern-0.10em b}\kern 0.05em{}} 
\def\nbar{\kern 0.10em \overline{\kern-0.10em n}\kern 0.05em{}} 
\def\pbar{\kern 0.10em \overline{\kern-0.10em p}\kern 0.05em{}} 
\def\Nbar{\kern 0.25em \overline{\kern-0.25em N}\kern 0.05em{}}

\def\BzBzb   {\ensuremath{\Bz {\kern -0.16em \Bzb}}\xspace}

\def\support{\footnote{Speaker on behalf of the {\babar} Collaboration}}

\def\Title#1{\begin{center} {\Large {\bf #1} } \end{center}}

\def\beq{\begin{equation}}
\def\eeq#1{\label{#1}\end{equation}}
\def\eeqn{\end{equation}}

\def\beqa{\begin{eqnarray}}
\def\eeqa#1{\label{#1}\end{eqnarray}}
\def\eeqan{\end{eqnarray}}

\let\bar=\overbar

\def\Dslash{\not{\hbox{\kern-4pt $D$}}}
\def\dslash{\not{\hbox{\kern-2pt $\del$}}}

\def\BR{\mbox{\rm BR}}

\def\msb{{\bar{\ssstyle M \kern -1pt S}}}

\begin{document}

\Title{{\Babar} Results For {\boldmath $\alpha$}:\\Measurement of  {\boldmath $\CP$}-Violating Asymmetries in {\boldmath $B^0 \to \left( \rho \pi \right)^0$} Using a Time-Dependent Dalitz Plot Analysis}

\bigskip\bigskip

\begin{raggedright}  

{\it Tomonari S. Miyashita\support\index{Miyashita, T. S.}\\
Department of Physics\\
Stanford University\\
Stanford CA, USA}\\
\bigskip\bigskip
Proceedings of CKM 2012, the 7th International Workshop on the CKM Unitarity Triangle, University of Cincinnati, USA, 28 September - 2 October 2012 
\end{raggedright}

\section{Introduction}

The decay $B^0 \rightarrow \pi^+ \pi^- \pi^0$ \footnote{Throughout this paper, whenever a mode is given, the charge conjugate is also implied unless indicated otherwise.} is well suited to the study of $\CP$ violation and has been previously explored by both the {\Babar}~\cite{ref:matt2007} and Belle~\cite{ref:belle2007} collaborations. Early studies of this mode involved ``quasi-two-body'' (Q2B) analyses that treated each $\rho$ resonance separately in the decays $B^0\rightarrow\rho^0\pi^0(\rho^0\rightarrow\pi^+\pi^-)$ and $B^0\rightarrow\rho^\pm\pi^\mp(\rho^\pm\rightarrow\pi^\pm\pi^0)$. However, as first pointed out by Snyder and Quinn~\cite{ref:snyder}, the use of a full time-dependent Dalitz plot (DP) analysis allows sensitivity to the interference effects caused by the relative strong and weak phases in the regions where the $\rho^+$, $\rho^-$, and $\rho^0$ resonances overlap.
This feature makes possible the unambiguous extraction of the strong and weak relative phases, and therefore the $\CP$-violating parameter $\alpha \equiv {\rm arg}[-V^{\phantom *}_{td}V^{*}_{tb}/(V^{\phantom *}_{ud}V^{*}_{ub})]$, where $V_{q q^\prime}$ are components of the Cabibbo-Kobayashi-Maskawa (CKM) quark-mixing matrix. A precision measurement of $\alpha$ is of interest because it serves to further test the standard model and constrain new physics that may contribute to loops in diagrams.

In this paper, we summarize an extensive reoptimization of an earlier {\Babar} analysis. We use the full ``on-resonance'' {\Babar} dataset of approximately $431$ ${\rm fb}^{-1}$ collected at the $\Upsilon(4S)$ resonance (an increase of 25\% in the number of $B$ decays) and include a number of improvements to both reconstruction and selection. Among these are improved charged-particle tracking, improved particle identification (PID), and a reoptimized multivariate discriminator (used both for event selection and as a variable in the final fit).

\section{Reconstruction and Event Selection}

The data used in this analysis were collected with the {\babar} detector at the PEP-II asymmetric-energy $e^+e^-$ storage ring at SLAC. Collisions occur at the $\Upsilon(4S)$ resonance energy (${\sqrt s} = 10.58\gevcc$), which frequently decays to $B{\overline B}$ pairs. We fully reconstruct the decay of one $B$ ($B_{3\pi}$) and use the decay of the other $B$ ($B_{\rm tag}$) to determine the flavor of $B_{\rm tag}$ at the time of its decay. Due to the asymmetric energies of the $e^+$ and $e^-$ beams, the $e^+e^-$ center-of-mass (CM) has a boost of $\beta\gamma \approx 0.56$ in the laboratory frame. The time-dependence of our analysis is measured using the distance along the beam axis between the $B_{3\pi}$ and $B_{\rm tag}$ decay vertices to calculate the time $\Delta t$ between the two decays.

Pairs of oppositely charged tracks are combined with $\pi^0 \to \gamma \gamma$ candidates to construct $B^0\rightarrow\pi^+\pi^-\pi^0$ candidates. The kinematics of $B$ meson decays that are fully reconstructed at {\babar} can be characterized by two variables: $m_{\rm ES}$ and $\Delta E$. The beam-energy-substituted mass $m_{\rm ES}$ is the invariant mass of the reconstructed $B$ candidate calculated under the assumption that its energy in the $e^+e^-$ CM frame is half the total beam energy. We define $m_{\rm ES} = \sqrt{\left[(s/2 + {\vec p_i} \cdot {\vec p_B})/E_i\right]^2 - |{\vec p_{B}}|^2}$, where $\sqrt{s}$ is the total beam energy in the $e^+e^-$ CM frame, $(E_i,{\vec p}_i)$ is the four-momentum of the $e^+e^-$ system in the laboratory frame, and ${\vec p}_B$ is the $B$-candidate momentum in the laboratory frame. The second kinematic variable is defined by $\Delta E = E^{*}_{B} - \frac{1}{2}\sqrt{s}$, where $E^{*}_{B}$ is the measured energy of the $B$ candidate in the $e^+e^-$ CM frame. We apply loose selection criteria using these variables and include them as inputs to the final fit.

Basic selection criteria are applied using quantities such as photon lateral moments (for the $\pi^0\rightarrow\gamma\gamma$ candidate), energy deposits, and track geometry parameters. Additionally, we use PID information to require that the $\pi^{\pm}$ candidates be consistent with the pion hypothesis. We apply a loose selection criterion using $\Delta t$ and include it as a variable in the final fit.

A further selection criterion is applied using a multivariate neural network (NN) discriminator. The discriminator serves to distinguish signal-like events (which have a more spherical topology) from $q{\overline q}$ continuum ($q=u,d,s,c$) background events which have a more collimated event shape. We train the discriminator using signal $B^0\rightarrow\rho\pi$ Monte Carlo (MC) and data collected below the $B^0{\overline B}^0$ threshold (to represent continuum background). A loose selection criterion is applied using the NN and we include it as a variable in our final fit.

While one would typically parameterize a Dalitz plot using the squared invariant masses of two pairs of daughter particles, practical considerations lead us to use a square Dalitz plot (SDP) parameterization in which the kinematically allowed region of DP phase space is mapped onto a unit square. The square Dalitz plot coordinates are $m^\prime$, which depends on $\pi^+\pi^-$ invariant mass, and $\theta^\prime$, which depends on the $\rho^0$ helicity angle $\theta_0$.

\section{Maximum Likelihood Fit}

We perform an unbinned extended maximum likelihood fit in order to extract event yields and physics parameters. The input variables are $m_{\rm ES}$, $\Delta E$, the NN output, and the three time-dependent-SDP variables $m^{\prime}$, $\theta^{\prime}$, and $\Delta t$. We also use $\sigma_{\Delta t}$ (the per-event uncertainty on $\Delta t$) as a scale factor in the signal $\Delta t$ resolution function. The likelihood function used in the fit consists of separate components for signal, continuum background, charged $B$ backgrounds, and neutral $B$ backgrounds. The signal component is subdivided into correctly reconstructed and misreconstructed components.

A probability density function (PDF) is associated with the distribution of each fit variable in each component of the likelihood function. Fixed and initial parameter values for these PDFs are obtained from fits to fully simulated MC (for signal $B$ background components) and either data collected below the $B^0{\overline B^0}$ threshold or a lower sideband in $m_{\rm ES}$ (for continuum).

We parameterize our signal PDF using 27 real-valued $U$ and $I$ coefficients, defined in terms of $B^0\rightarrow\rho^{+,-,0}\pi^{-,+,0}$ and ${\overline B}^0\rightarrow\rho^{+,-,0}\pi^{-,+,0}$ decay amplitudes ($A^{+,-,0}$ and ${\overline A}^{+,-,0}$, respectively) as $U_{\kappa}^{\pm} = |A^\kappa|^2 \pm |{\overline A}^\kappa|^2$, 
$U_{\kappa \sigma}^{\pm, {\rm Re(Im)}} = {\rm Re(Im)} \left[ A^\kappa A^{\sigma *} \pm {\overline A}^\kappa {\overline A}^{\sigma *} \right]$, 
$I_\kappa = {\rm Im} \left[ {\overline A}^\kappa A^{\kappa *} \right] ,\quad I_{\kappa \sigma}^{\rm Re} = {\rm Re} \left[ {\overline A}^\kappa A^{\sigma *} - {\overline A}^\sigma A^{\kappa *} \right]$, and 
$I_{\kappa \sigma}^{\rm Im} = {\rm Im} \left[ {\overline A}^\kappa A^{\sigma *} + {\overline A}^\sigma A^{\kappa *} \right]$
where $\kappa\in(+,-,0)$ and $\kappa\sigma\in(+-,+0,-0)$.
\noindent These coefficients provide an alternative parameterization to tree and penguin amplitudes (as well as $\alpha$) or to the amplitudes $A^\kappa$ and ${\overline A}^\kappa$~\cite{ref:snyderquinnusis}. The $U$ and $I$ parameters can also be directly related to the Q2B parameters (${\mathcal C}$, $\mathcal S$, $\Delta {\mathcal C}$, $\Delta {\mathcal S}$, ${\mathcal A}_{\rho\pi}$, ${\mathcal C}_{00}$, ${\mathcal S_{00}}$, and $f_{00}$) often used in $\CP$-violation analyses~\cite{ref:cpv}.

We include the $\rho(1450)$ in the final fit with an assumption that the relative magnitudes and phases between the three $\rho(1450)$ resonances are the same as for the $\rho(770)$. Whereas there is reasonable motivation for this assumption in the case of the $\rho(1450)$ since the $\rho(770)$ and $\rho(1450)$ have the same quantum numbers, the $\rho(1700)$ does not share these quantum numbers ($\ell=2$ instead of $0$). Since the $\rho(1700)$ is not expected to have a large contribution to the decay rate, we excluded the $\rho(1700)$ from the fit and associate a systematic uncertainty with this omission. We also associate systematic uncertainties with the expected numbers of $B$ background events (which are fixed in the fit), the masses and widths used in our $\rho$ lineshapes, possible contributions from uniform backgrounds, and various other small contributions. While the most significant systematic uncertainty is that associated with the exclusion of the $\rho(1700)$ from the nominal fit model, even this contribution is found to be small and the sensitivity remains dominated by statistical uncertainties.

\section{Results}
\label{sec:resultssec}

\begin{figure*}[!htb]
\begin{center}
\includegraphics[clip=true,trim=25 370 25 80,width=0.99\textwidth]{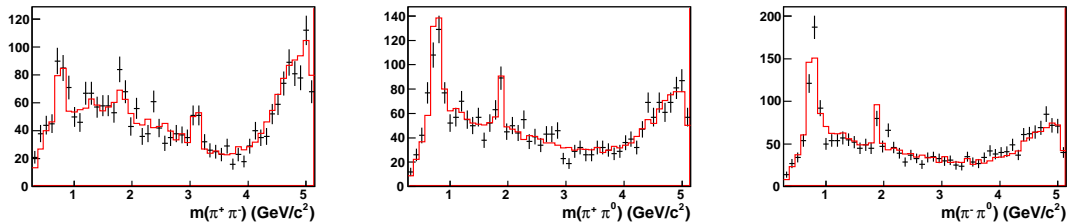}
\caption{Overlay plots of on-resonance data (points with error bars) and parameterized MC generated from the final fit results (red line) with 10 times the number of events in data. The MC histograms are scaled to have the same integral as the data histograms onto which they are overlaid. A tight cut is applied to the NN output to enhance the signal component. }
\label{fig:overlayFigureEnhanced}
\end{center}
\end{figure*}

From an on-resonance dataset containing 53,084 candidates, the fit extracts 2,940$\pm100$ signal events and 46,750$ \pm 220$ continuum events. Figure~\ref{fig:overlayFigureEnhanced} contains overlaid $\pi\pi$ invariant mass plots of the data used in the final fit and parameterized MC generated using the results of the final fit. A study of the $U$ and $I$ parameters (see Sec.~\ref{sec:robustness}) exhibits negligible bias in their extraction, and good robustness in the presence of statistical fluctuations.

The complete final results from the extended maximum likelihood fit, including statistical and systematic uncertainties, are provided in the Physical Review D article associated with this analysis (currently in preparation). We find the extracted $U$ and $I$ parameter values as well as the extracted Q2B parameter values to be consistent with the previous results from {\babar} and Belle. The sensitivity of the present analysis is much improved over the previous {\babar} analysis, with an average ratio of statistical uncertainties on $U$ and $I$ parameters relative to the previous analysis of 0.47. This is a larger increase in sensitivity than can be explained by the $25\%$ increase in the size of the data sample and it may be attributed to the many improvements made in this analysis. Similarly, the average ratio of the statistical uncertainty on the Q2B parameters in this analysis relative to the previous {\babar} analysis is $0.61$. The final fit results for the Q2B parameters are provided in Table~\ref{tab:q2b}. A study of the Q2B parameters (see Sec.~\ref{sec:robustness}) exhibits negligible bias in their extraction, and good robustness in the presence of statistical fluctuations.

\begin{figure}
\begin{center}
\includegraphics[width=0.45\textwidth]{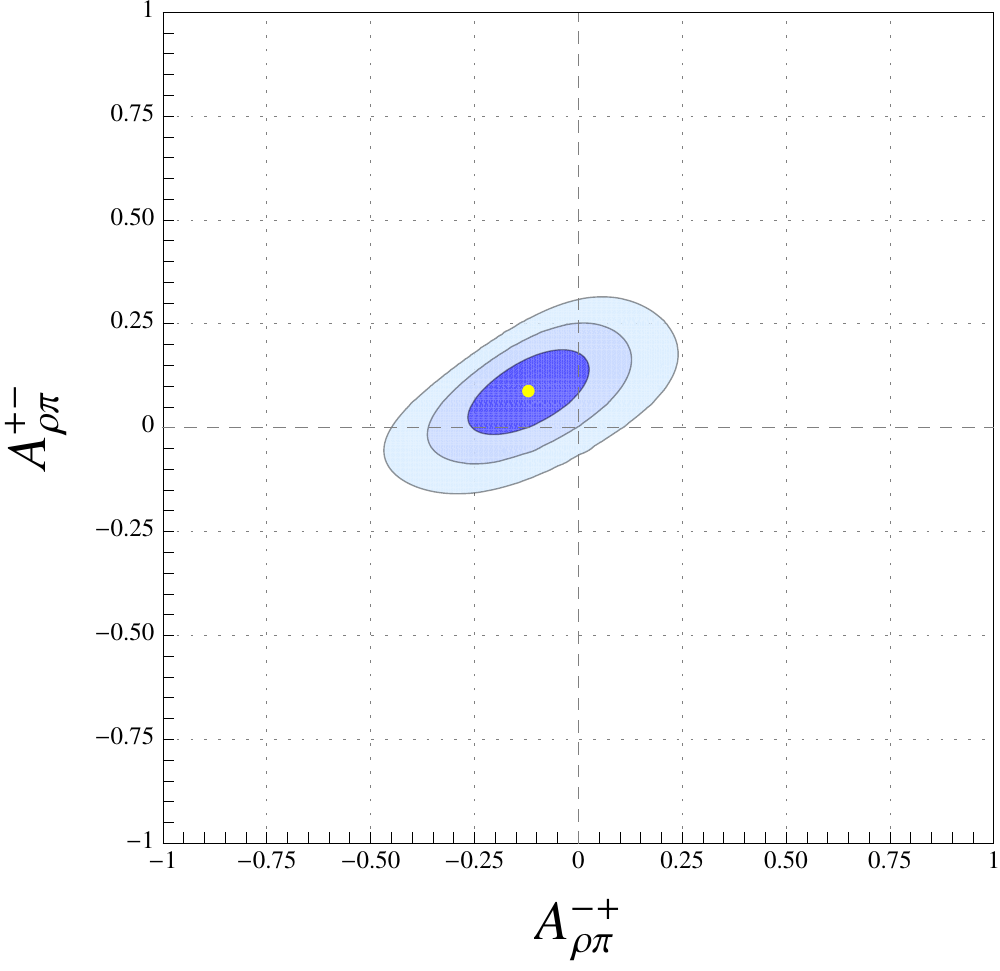}
\caption{Combined statistical and systematic two-dimensional likelihood scan of  ${\mathcal A}^{+-}_{\rho\pi}$ vs.~${\mathcal A}^{-+}_{\rho\pi}$ with $68.3\%$, $95.5\%$, and $99.7\%$ confidence-level contours ($\Delta\chi^2=\{2.30,6.18,11.83\}$). The yellow point indicates the central value.}
\label{fig:contours}
\end{center}
\end{figure}

The parameters ${\mathcal A}_{\rho\pi}$ and ${\mathcal C}$ can be transformed into the direct $\CP$-violation parameters ${\mathcal A}^{+-}_{\rho\pi}$ and ${\mathcal A}^{-+}_{\rho\pi}$ as described in Ref.~\cite{ref:matt2007}.
We extract the central values and uncertainties for these parameters using a $\chi^2$ minimization in the two-dimensional plane corresponding to ${\mathcal A}^{+-}_{\rho\pi}$ vs. ${\mathcal A}^{-+}_{\rho\pi}$.
From this two-dimensional scan, we find ${\mathcal A}^{+-}_{\rho\pi} = {\phantom -} 0.09 ^{+0.05}_{-0.06} \pm 0.04$ and ${\mathcal A}^{-+}_{\rho\pi} = -0.12 \pm 0.08 ^{+0.04}_{-0.05}$. A plot of this scan is provided in Fig.~\ref{fig:contours}. The origin corresponds to no direct $CP$ violation and lies on the $96.0\%$ confidence level contour ($\Delta \chi^2=6.42$). The corresponding $p$ value for the hypothesis of no direct $CP$ violation is $4.0\%$.

\begin{table}

\begin{multicols}{2}
\begin{center}
\begin{tablehere}
\begin{tabular}{ l | l  l l }
Param       &     Value        	    &   $\sigma_{\rm stat}$   &  $\sigma_{\rm syst}$ \\ \hline
${\mathcal A}_{\rho \pi}   $   &                $-0.100$      &   0.029         &  0.021       \\
${\mathcal C}              $   &     ${\phantom -}0.016$      &   0.059	      &  0.036	     \\
$\Delta{\mathcal C}        $   &     ${\phantom -}0.234$      &   0.061	      &  0.048	     \\
${\mathcal S}              $   &     ${\phantom -}0.053$      &   0.081 	      &  0.034	     
\end{tabular}
\end{tablehere}

\begin{tablehere}
\begin{tabular}{ l | l  l l }
Param       &     Value        	    &   $\sigma_{\rm stat}$   &  $\sigma_{\rm syst}$ \\ \hline
$\Delta{\mathcal S}        $   &     ${\phantom -}0.054$      &   0.082	      &  0.039	     \\
${\mathcal C}_{00}                    $   &     ${\phantom -}0.19$       &   0.23	      &  0.15	     \\
${\mathcal S}_{00}                    $   &                $-0.37$       &   0.34	      &  0.20	     \\
$f_{00}                    $   &     ${\phantom -}0.092$      &   0.011         &  0.008       
\end{tabular}
\end{tablehere}

\end{center}
\end{multicols}

\caption{Quasi-two-body parameter values and uncertainties corresponding to the fit to the complete on-resonance dataset. }
\label{tab:q2b}
\end{table}

\subsection{\boldmath $\alpha$ Scan Results}
\label{sec:scan}

In order to extract likely values of $\alpha$ from the $U$ and $I$ parameters obtained in our final fit, we perform a scan of $\alpha$ from $0^\circ$ to $180^\circ$. At each scan point, a $\chi^2$-minimization fit is performed using the combined statistical and systematic covariance matrices from our nominal fit. As the scan proceeds, a minimum $\chi^2$ value is extracted from the fit at each value of $\alpha$. We convert these $\chi^2$ values to ``$\Sigma$'' values by calculating the $\chi^2$ probability of each value according to $\Sigma \equiv \int_a^\infty f(x;1) dx$, where $a$ is the difference between the $\chi^2$ at the current scan point and the minimum $\chi^2$ for all the scan points, and $f(x;1)$ is a $\chi^2$ distribution with one degree of freedom. The variable ``$\Sigma$'' corresponds to what is commonly referred to as ``1$-$Confidence Level'' (1$-$C.L.).

Following the methods employed in Belle's 2007 $B^0\rightarrow\rho\pi$ analysis~\cite{ref:belle2007} and described in~\cite{ref:snyderquinnusis}, we perform a further $\alpha$ scan that makes use of measurements from the charged decays $B^\pm\rightarrow\rho^{\pm,0}\pi^{0,\pm}$. Amplitudes for these modes can be related to amplitudes in the neutral $B$ modes due to isospin relations. These relations result in four constraint equations while introducing only two new free parameters in the fit (which arise from the unknown relative magnitude and phase of the charged-$B$ and neutral-$B$ decay amplitudes).

\begin{figure}[!htb]
		\includegraphics[width=0.49\textwidth]{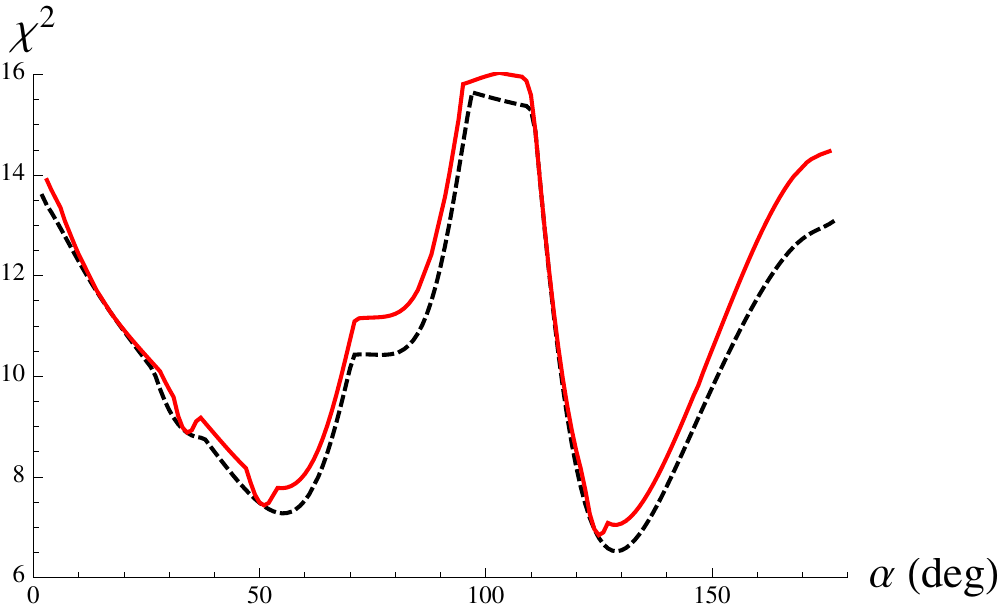}
		\includegraphics[width=0.49\textwidth]{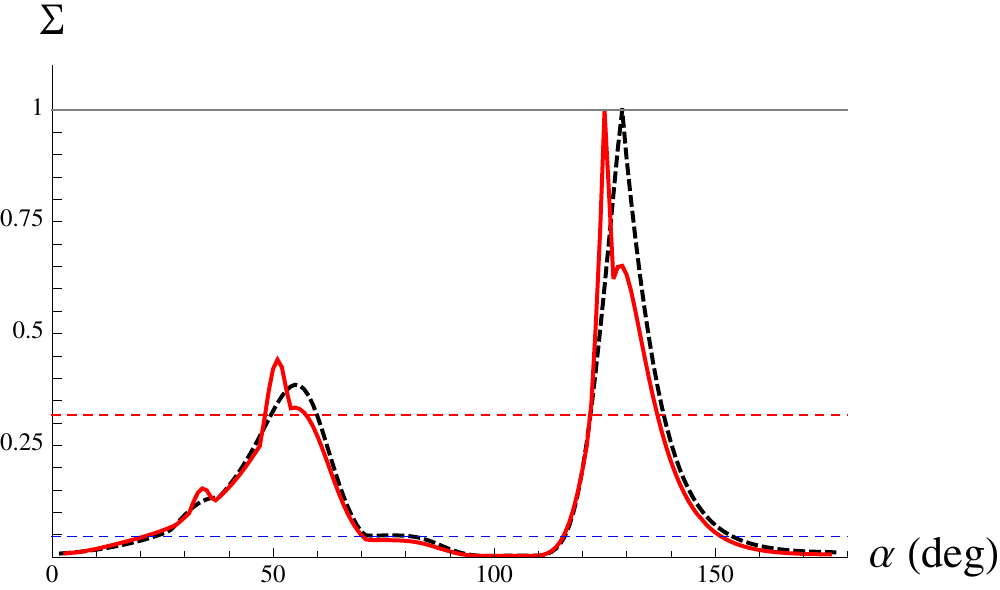}
		\caption{Left: Scans of minimum $\chi^2$ values as a function of $\alpha$. Note that the origin on the vertical scale is suppressed. Right: Scans of $\Sigma$ values as a function of $\alpha$. The upper and lower horizontal dashed lines correspond to $\Sigma=0.05$ and $0.32$, respectively. All scans are based on the fit to the full on-resonance dataset and include contributions from both statistical and systematic uncertainties.}
		\label{fig:chi}
\end{figure}

Graphs of the $\chi^2$ values from our final $\alpha$ scans with isospin constraints (solid red) and without isospin constraints (dashed black) are provided in the left plot of Fig.~\ref{fig:chi}. The corresponding $\Sigma$ distributions are given in the right plot of Fig.~\ref{fig:chi}. Importantly, our robustness studies (see Sec.~\ref{sec:robustness}) indicate that the $\Sigma$ scan is not robust with our current sample size (or those available to the previous {\babar} and Belle analyses) and cannot be interpreted in terms of Gaussian statistics. 

\section{Robustness Studies}
\label{sec:robustness}

An important component of this analysis is a set of studies which assess the robustness with which the fit framework extracts statistically accurate values and uncertainties for the $U$ and $I$ parameters, the Q2B parameters, and $\alpha$ by employing 25 MC simulated samples generated with a parameterized detector simulation and with signal and background contributions corresponding to those expected in the full on-resonance dataset. The samples are simulated using physical $U$ and $I$ parameters generated based on specific tree and penguin amplitudes and $\alpha=89^\circ$ (approximately the world average). Each simulated dataset is generated with the same parameter values, but a different random-number seed. By examining the results of fits to each of these simulated datasets, we assess the robustness of the fits. Comparisons of the extracted $U$ and $I$ and Q2B parameter values with the generated values find that all parameters are robustly extracted with negligible bias and well estimated uncertainties. More significant are the results of the $\alpha$ robustness study.

\begin{figure}[!htb]
\begin{center}
\includegraphics[width=0.48\textwidth]{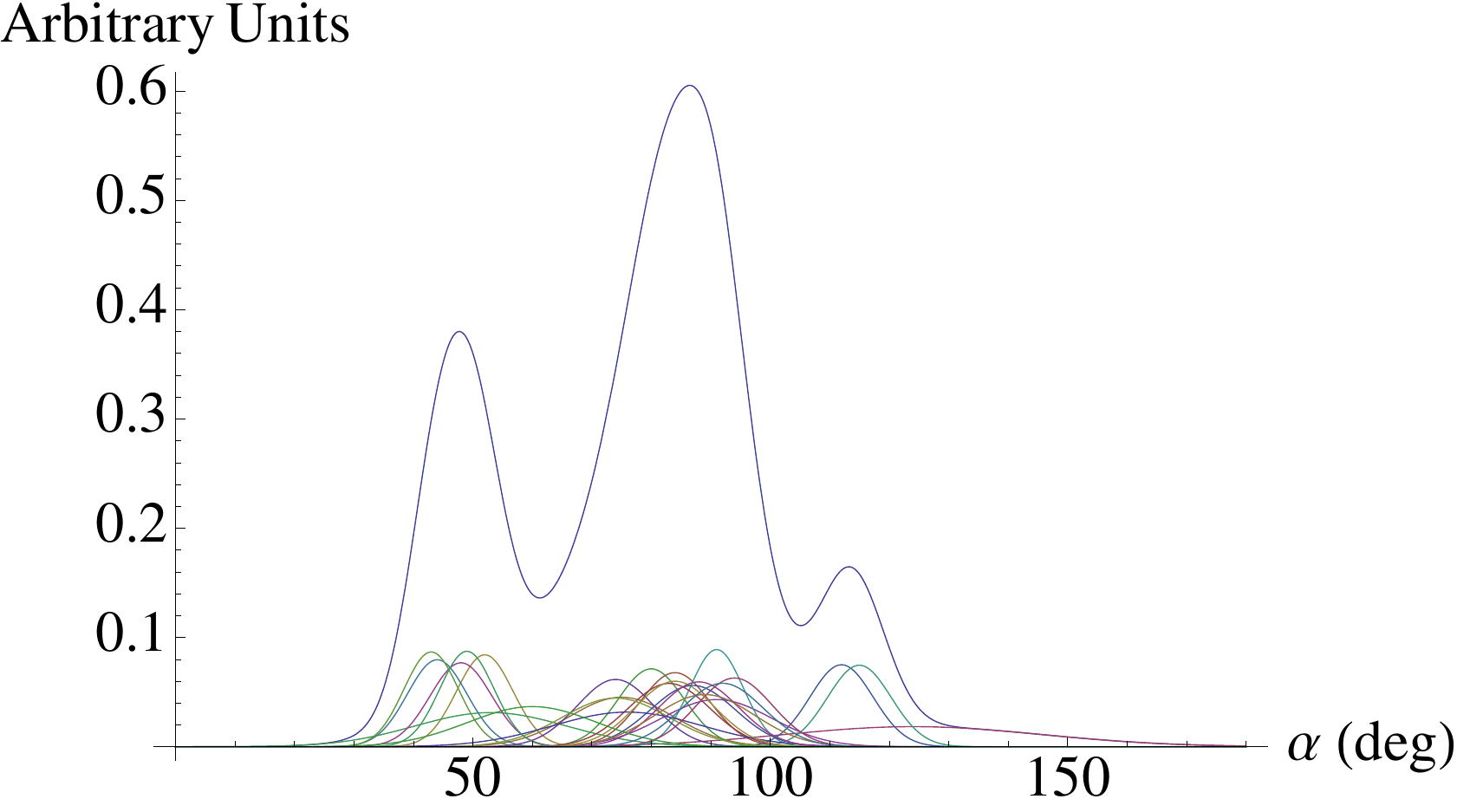}
\includegraphics[width=0.48\textwidth]{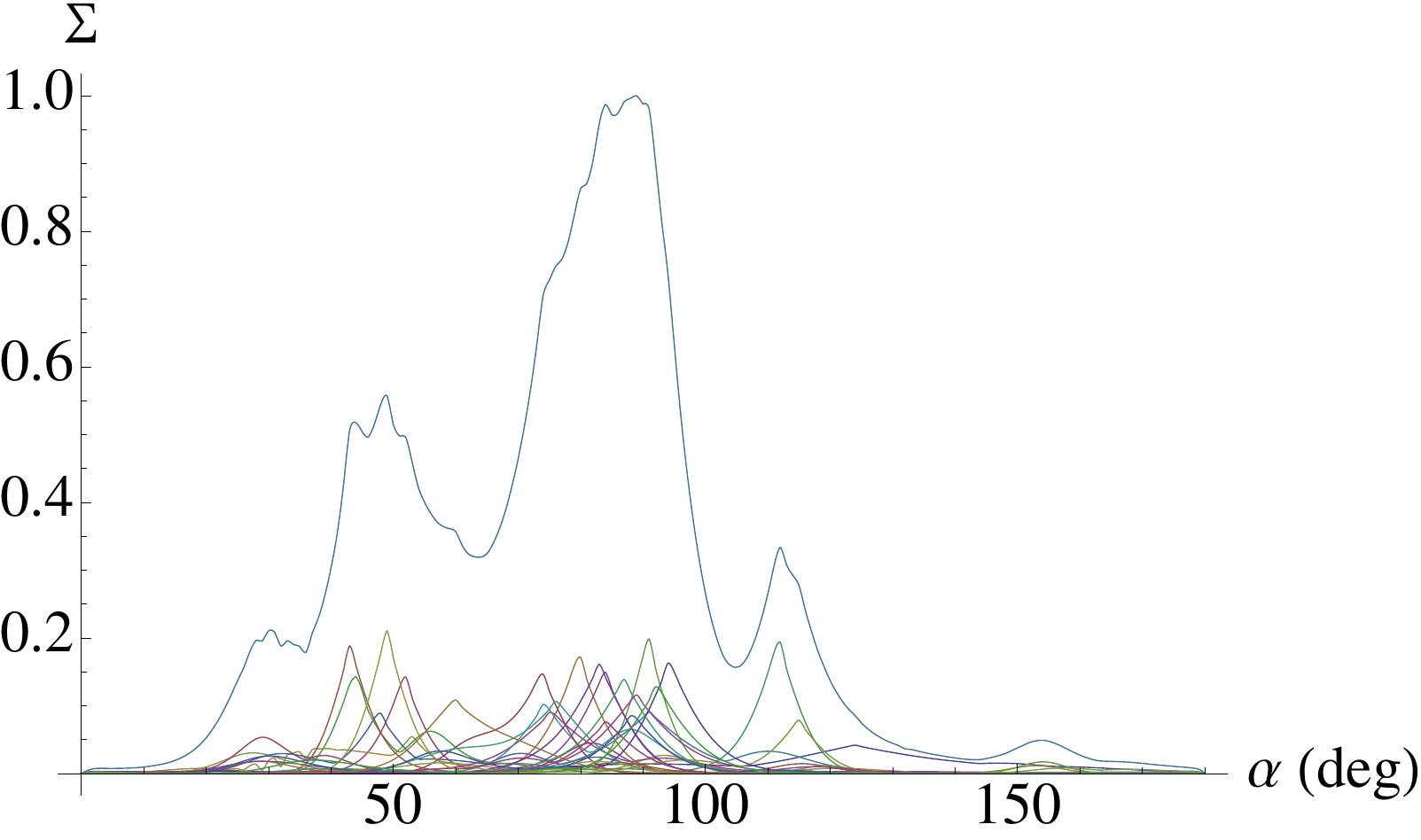}
\caption{Left: Plot of the sum of 25 normalized Gaussians based on peaks from scans of parameterized MC generated with $\alpha=89^\circ$. Also plotted are the individual Gaussians which contribute to the total PDF. Right: Plot of the total $\Sigma$ distribution obtained by summing up all 25 $\Sigma$ scans of parameterized MC generated after normalizing their areas to 1. Also plotted are the individual $\Sigma$ scans.}
\label{fig:gaussianRobustness}
\end{center}
\end{figure}

A one-dimensional likelihood scan of $\alpha$ is performed using the results of the fits to each of the 25 MC samples. For 8 of the 25 scans, the extracted value of $\alpha$ lies more than $3\sigma$ from the generated value. Examining the individual $\alpha$ scans reveals three distinct solutions for $\alpha$ that tend to be favored (including the generated value of $89^\circ$) and each scan tends to include at least one secondary peak in addition to the primary peak. The left plot in Figure~\ref{fig:gaussianRobustness} illustrates the three solutions for $\alpha$ by providing the sum of 25 normalized Gaussians with means and widths determined by the peak positions and symmetric errors extracted from the 25 $\alpha$ scans. Because the errors are not truly Gaussian, the plot provides an incomplete picture of the scan results. A better illustration is provided by the right plot in Figure~\ref{fig:gaussianRobustness}, which displays the total $\Sigma$ distribution obtained by summing all 25 $\alpha$ scans after normalizing each to the same area. The total distribution is scaled so that it peaks at 1. The final PDF closely resembles that obtained by naively summing Gaussian distributions, though it exhibits more fine features. Again, the distribution indicates three distinct solutions for $\alpha$, with the generated value of $89^\circ$ being favored. At the $1\sigma$ level ($\Sigma$=0.32), the total scan distribution allows both the central and left peak. The presence of these secondary solutions indicates that with the current signal sample size and background levels, there is still a significant possibility that the favored value of $\alpha$ in a particular scan will correspond to a secondary solution. 

\section{Conclusions}
\label{sec:conclusion}

We have performed a time-dependent Dalitz plot analysis of the mode $B^0 \rightarrow (\rho \pi)^0$ in which we extract 26 $U$ and $I$ parameter values describing the physics involved, as well as their full statistical and systematic covariance matrices in an extended unbinned maximum likelihood fit to the full {\babar} dataset. From these fit results, we extract standard quasi-two-body parameters with values given in Table~\ref{tab:q2b}. These Q2B values are consistent with the results of the 2007 {\babar}~\cite{ref:matt2007} and Belle~\cite{ref:belle2007} analyses, but exhibit significantly increased sensitivity. We also perform a two-dimensional likelihood scan of the direct $\CP$-violation asymmetry parameters for $B^0\rightarrow\rho^\pm\pi^\mp$ decays, finding the change in $\chi^2$ between the minimum and the origin (corresponding to no direct $\CP$-violation) to be $\Delta \chi^2=6.42$ (approximately $2\sigma$). Finally, we perform one-dimensional likelihood-scans of the unitarity angle $\alpha$ both with and without isospin constraints from other modes (see Fig.~\ref{fig:chi}).

Notably, we also perform a series of robustness studies in order to determine how reliably our fit framework extracts the actual value of physics parameters. The studies reveal that with our current signal sample size and background suppression, we can reliably extract the $U$ and $I$ parameters as well as the Q2B parameters, but the extraction of $\alpha$ in $B^0\rightarrow\rho\pi$ is {\em not} statistically robust. This result has consequences not only for this analysis, but earlier {\babar} and Belle published results as well. Namely, it calls into question the reliability of the $\alpha$ values quoted in previous analyses and indicates that with the current sensitivity, we should not treat the $\alpha$ scan as a straightforward measurement of $\alpha$. This analysis would benefit greatly from increased sample sizes available at high-luminosity experiments such as Belle II.

\end{document}